\documentclass[aps,prd,reprint,showpacs,floatfix]{revtex4-1}

\usepackage[english]{babel}
\usepackage{graphicx}
\usepackage{array}
\usepackage{dcolumn}
\usepackage{bm}
\usepackage{amssymb}
\usepackage{amsmath}
\usepackage{amsfonts}
\usepackage{amsbsy}

\begin{document}

\title{Reanalysis of nuclear spin matrix elements for
dark matter spin-dependent scattering }

\author{M.~Cannoni}

\affiliation{Departamento de F\'isica Aplicada, Facultad de Ciencias
Experimentales, Universidad de Huelva, 21071 Huelva, Spain}

\begin{abstract}

We show how to include in the existing calculations for nuclei other than $^{129}$Xe, $^{131}$Xe,
the corrections to the isovector coupling arising in chiral effective field theory recently found in Ref.~\cite{Menendez1}. 
The dominant, momentum independent, 2-body currents effect can be taken into account by formally redefining the 
static spin matrix elements $\langle \mathbf{S}_{p,n} \rangle$. By further using the normalized form factor 
at $q\neq 0$ built with the 1-body level structure functions, we show that the WIMP-nucleus cross section and the 
upper limits on the WIMP-nucleon cross sections
coincide with the ones derived using the exact functions at the 2-body level.
We explicitly show it in the case of XENON100 limits on the WIMP-neutron cross section 
and we recalculate the limits on the WIMP-proton spin dependent cross section 
set by COUPP. 
We also give practical formulas to obtain $\langle \mathbf{S}_{p,n} \rangle$ given the structure functions in the 
various formalisms/notations existing in literature. 
We argue that the standard treatment of the spin-dependent cross 
section in terms of three independent isospin functions, $S_{00}(q)$, $S_{11}(q)$, $S_{01}(q)$, is redundant in the 
sense that the interference function $S_{01}(q)$ is the double product 
$|S_{01}(q)|=2\sqrt{S_{00}(q)}\sqrt{S_{11}(q)}$ even when including the new effective field theory corrections.

\end{abstract}

\pacs{95.35.+d, 12.60.Jv}
	

\maketitle

\section{Introduction}
\label{Sec:intro}

The formalism for the cross section of the elastic scattering of weakly interacting massive 
particles (WIMP) off nuclei, was formulated many years ago for the neutralino~\cite{engel,engelrev}, a well motivated 
supersymmetric dark matter candidate~\cite{jungman}. 
The dominant interactions of a non-relativistic WIMP with a nucleus
are the coherent spin-independent and spin-spin interactions~\cite{goodman,kurylov}. 

Nowadays many other particle physics models
that predict  a dark matter candidate have been proposed. However, in the non--relativistic limit,
only a limited number of operators remain, as has been shown with the help of model independent effective 
field theory~\cite{Fitzpatrick1}. To calculate the contribution to the cross section of these operators, the 
corresponding nuclear responses must be evaluated.  

The scope of this paper is to clarify some previously unrecognized properties of the response functions of the spin 
dependent formalism and show how to take into account the effects of low energy pion physics recently 
found in Ref.~\cite{Menendez1} and included in a new calculation of the structure functions of $^{129}$Xe
and $^{131}$Xe.  

In Section~\ref{Sec:1} and Appendix~\ref{App:A} we show that the isospin interference function $S_{01}$
is not an independent function but that it is determined by the isoscalar and isovector 
functions $S_{00}$ and $S_{11}$. 

In Section~\ref{Sec:2} we show how to account for the strong interactions 
correction of Ref.~\cite{Menendez1} in a general way.
We also compare, with the aid of Appendix~\ref{App:B},
the results of different calculations for nuclei of experimental interest.

In Section~\ref{Sec:3} we show explicitly, using the last data of XENON100~\cite{xenon100_225_SD,xenon100_225_SI}, 
that our method furnishes the same upper limit on the WIMP-neutron cross section as the
one obtained with the exact spin structure functions. 
To illustrate the impact of these effects we recalculate the limits on the WIMP-proton cross 
section using the last data of the COUPP experiment~\cite{coupp}.

A brief summary of the main results of the paper is given in Section~\ref{Sec:4}.

\section{Only two functions}
\label{Sec:1}

The effective axial interaction of a spin 1/2 Majorana fermion with nucleons
\[
\mathcal{L}=2\sqrt{2}{G_F} a_{N}\overline{\chi}\gamma^\mu \gamma^5 {\chi}
\overline{N}\gamma_\mu \gamma^5 N
\] 
reduces in the non-relativistic limit to the operator $(2\sqrt{2}{G_F})a_N 4m_\chi m_N 
(4\mathbf{S}_\chi \cdot \mathbf{S}_N)$
and leads to a WIMP-nucleus cross section that, after Refs.~\cite{engel,engelrev}, is given in the form
\begin{eqnarray}
\frac{d\sigma}{dq^2} &=&\frac{8G^2_F}{v^2}\frac{S(q)}{2J+1},\\
S(q) &=& a^2_0 S_{00}(q) +  a_0 a_1 S_{01}(q) + a^2_1 S_{11}(q).
\label{Eq:Sq}
\end{eqnarray}

Although nuclear physics calculations are carried out in the isospin representation defined by 
the isoscalar coupling $a_0=a_p + a_n$, and the isovector coupling $a_1 =a_p -a_n$, for practical 
purposes is natural to think in terms of neutrons and protons. 
The quantities usually employed by experimentalists are thus the couplings 
$a_{p}=(a_0 + a_1)/2$, 
$a_{n}=(a_0 - a_1)/2$,
and the spin matrix elements of the protons and neutrons group in the nuclear ground state with 
total angular momentum $J$:
$\langle \mathbf{S}_{p} \rangle \equiv  \langle J,M=J|\sum_{i=1}^{Z} S^z_{i}|J,M =J  \rangle$,
$\langle \mathbf{S}_{n} \rangle \equiv  \langle J,M=J|\sum_{i=1}^{A-Z} S^z_{i}|J,M =J  \rangle$.
In the protons-neutrons representation, Eq.~(\ref{Eq:Sq}) at zero momentum transfer reduces to 
\begin{equation}
S(0)=\frac{2J+1}{\pi}\frac{J+1}{J}|a_p \langle \mathbf{S}_{p} \rangle
+a_n \langle \mathbf{S}_{n} \rangle|^2.
\label{Eq:S0}
\end{equation}
The differential cross section is usually then rewritten as
\begin{equation}
\frac{d\sigma}{dE_R}  =\frac{m_A}{2 \mu^2_A v^2 } \sigma_A (0) \Phi^{SD} (q), 
\label{Eq:dsigmadE}
\end{equation}
where the total WIMP-nucleus cross sections at $q=0$ is 
\begin{equation}
\sigma_{{A}}(0) =8G^2_F \frac{\mu^2_A}{2J+1}S(0),
\label{Eq:sigmatot}
\end{equation}
and the form factor normalized to 1 at $q=0$
\begin{equation}
\Phi^{SD}(q)=\frac{S(q)}{S(0)}.
\label{Eq:FFSD}
\end{equation}

The functional form of the structure functions $S_{ij}$ obtained from shell-model calculations
is {\it exactly} a polynomial times an exponential in the dimensionless variable 
$y=(qb/2)^2$ (or $u=2y$), being $b$ the oscillator size parameter, 
if the single particle states used to build the 
nuclear wave function are wave functions of the three-dimensional harmonic oscillator potential.
This is not the case if the eigenfunctions of the Wood-Saxon potential are used~\cite{mica,dimitrov,resselldean}
and when the effects discussed in Section~\ref{Sec:2} are included;
anyway, it is always possible to {\it fit} them with a polynomial or an exponential times a
polynomial in $y$ or $u$. 
Hence in general
\begin{equation}
S_{ij}=e^{-2y}\sum^{k_{\text{max}}}_{k=0} c^{(k)}_{ij}y^k,\;\text{or }
S_{ij}=\sum^{k_{\text{max}}}_{k=0} c^{(k)}_{ij}y^k.
\label{Eq:Sijy}
\end{equation}

It thus seems that three nuclear response function in the isospin 
representation $S_{11}$, $S_{00}$, $S_{01}$ are necessary to furnish the cross section 
at finite momentum transfer, while only the two numbers $\langle \mathbf{S}_{p,n} \rangle $ 
are necessary at $q=0$. 
This asymmetry is unnatural, $q=0$ is just a value of the variable $q$, and actually it does not hold.

The functions $S_{00}(q)$, $S_{11}(q)$ are monotonic decreasing, always positive at every $q$,
while $S_{01}$ is always positive or always negative at every $q$.
The simple reason behind this behavior, as shown in Appendix~\ref{App:A}, is that we can define 
two functions $S_{0}(q)$ and $S_1(q)$ (that maintain their sign) such that 
the isoscalar and the isovector functions $S_{00}$ and $S_{11}$ are
nothing but the square of $S_0$ and $S_1$, 
\begin{equation}
S_{00}(q)\equiv S^2_0(q),\;\;\;\; S_{11}(q)\equiv S^2_1(q),
\label{Eq:S0S1}
\end{equation}
and the interference function is the double product, 
\begin{equation}
S_{01}\equiv 2S_0(q) S_1(q),\;\;\;\;|S_{01}(q)|= 2 \sqrt{S_{00}(q) S_{11}(q)},
\label{Eq:S01}
\end{equation}
with sign $\rho_{01}=\text{sgn}(S_{01})$ given by the relative sign between $S_0(q)$ and $S_1(q)$.

\begin{figure}[b!]
\includegraphics*[scale=0.59]{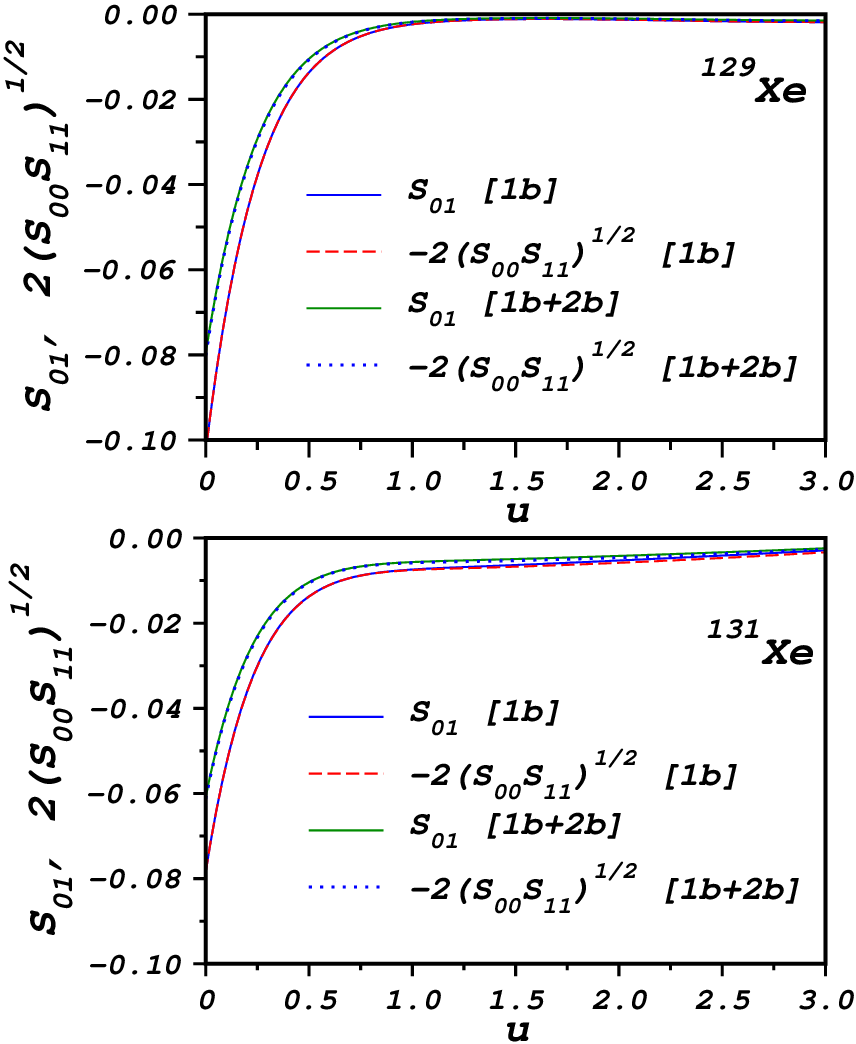}
\caption{The isoscalar-isovector interference function $S_{01}$ and the double product
of Eq.~(\ref{Eq:S01}) in the case of $^{129}$Xe and $^{131}$Xe from Ref.~\cite{Menendez1}. }
\label{Fig:1}
\end{figure}

\begin{figure}[t!]
\includegraphics*[scale=0.59]{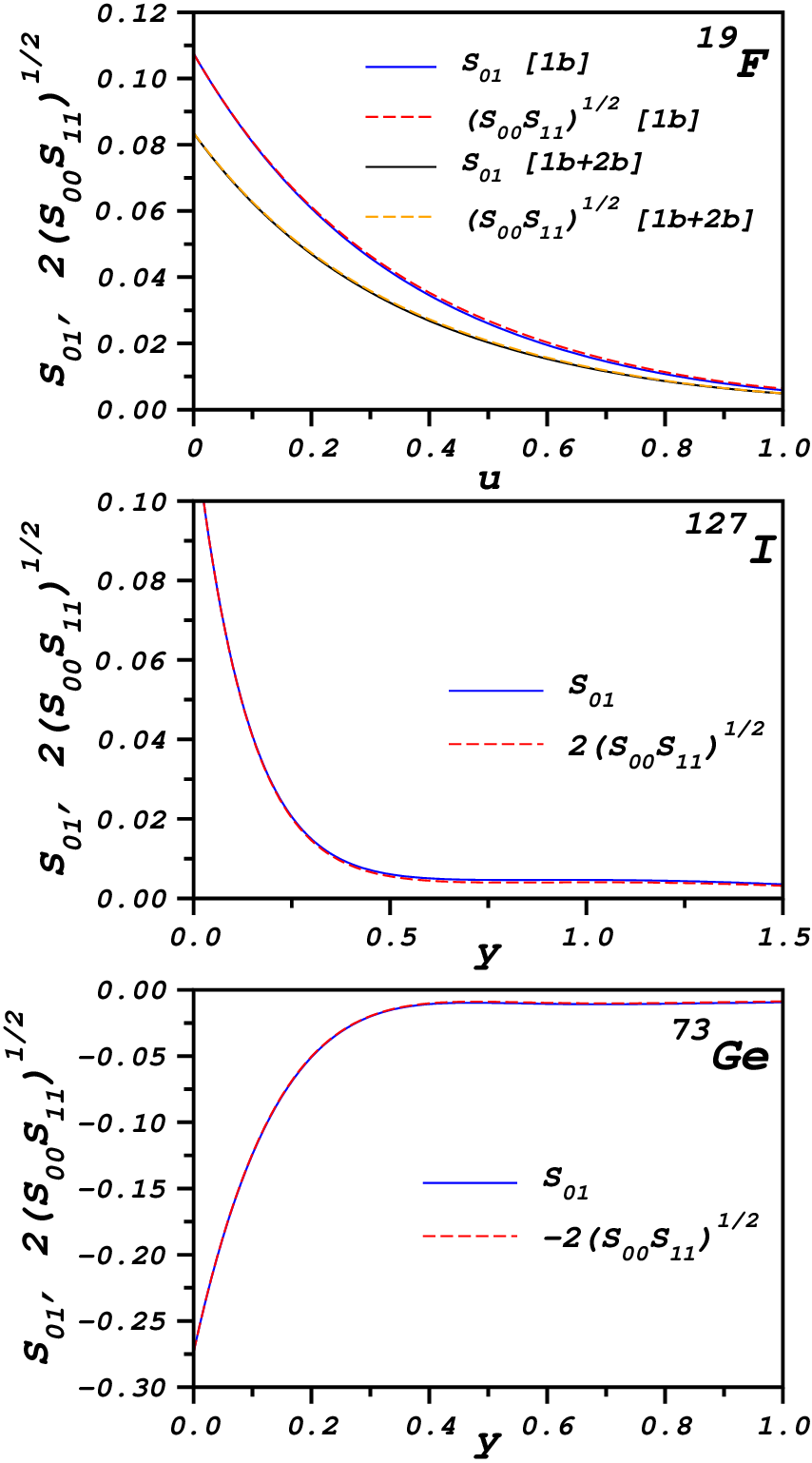}
\caption{The isoscalar-isovector interference function $S_{01}$ and the double product
of Eq.~(\ref{Eq:S01}). Functions for $^{19}$F are from Ref.~\cite{vergados_nuovo}: note that 
these authors use a different normalization, hence the factor 2 is missing. Functions 
for $^{73}$Ge are from Ref.~\cite{dimitrov}, for $^{127}$I from Ref.~\cite{resselldean} (Bonn A set). }
\label{Fig:2}
\end{figure}

We explicitly show this fact in Fig.~\ref{Fig:1} and Fig.~\ref{Fig:2} for some isotopes 
used in actual experiments. 
The functions for $^{129}$Xe and $^{131}$Xe are from 
Ref.~\cite{Menendez1}.
The functions referring to $^{73}$Ge are taken from 
Ref.~\cite{dimitrov}, the ones for $^{127}$I from Ref.~\cite{resselldean} (Bonn A potential set) 
and the ones for $^{19}$F are from Ref.~\cite{vergados_nuovo}: note that in this last 
case the authors use a different normalization, hence the factor 2 is missing.
See also Fig.~\ref{Fig:6} in Appendix~\ref{App:B}, bottom panel, for $^{23}$Na.
The small differences that one can find in some cases can be ascribed to the intrinsic
uncertainties of the numerical fits and numerical rounding.
The labels "[1b]" and "[1b+2b]" indicate that the calculations are done with 1-body operators,
"[1b+2b]" on the other hand means that the 2-body operators effects that will be discussed in the next section 
are included.

The physics at $q=0$
is determined by the coefficients of order zero of the polynomials in Eq.~(\ref{Eq:Sijy}),
\begin{equation}
S_{00}(0)=c^{(0)}_{00},\;S_{11}(0)=c^{(0)}_{11},\;\rho_{01}=\text{sgn}(c^{(0)}_{01}).
\label{Eq:coeff0}
\end{equation}
Comparing Eq.~(\ref{Eq:Sq}) evaluated at $q=0$ with Eq.~(\ref{Eq:S0}) we have 
\begin{equation}
\langle \bm{S}_{p,n} \rangle^2 = {\frac{\pi}{2J+1}\frac{J}{J+1}}({S_{00}(0)} + {S_{11}(0)}\pm S_{01}(0)  ),
\nonumber
\end{equation}
being the plus sign associated with the protons and the minus sign with the neutrons. 
The static spin matrix elements, in the light of Eqs.~(\ref{Eq:S0S1}),~(\ref{Eq:S01}),~(\ref{Eq:coeff0}),  
are then given by 
\begin{eqnarray}
\langle \bm{S}_{p} \rangle &=&
\lambda \sqrt{\frac{\pi}{2J+1}\frac{J}{J+1}}(\sqrt{c^{(0)}_{00}} + \rho_{01}\sqrt{c^{(0)}_{11}}  ),\nonumber\\
\langle \bm{S}_{n} \rangle &=&
\lambda \sqrt{\frac{\pi}{2J+1}\frac{J}{J+1}}(\sqrt{c^{(0)}_{11}} - \rho_{01}\sqrt{c^{(0)}_{11}}  ),
\label{Eq:+SpSn}
\end{eqnarray}
where $\lambda$ is + or - in both, depending on the given nucleus.
We will use these formulas when will discuss the spin matrix elements of $^{19}$F.

\section{Including strong interaction corrections }
\label{Sec:2}

In Ref.~\cite{Menendez1} it has been shown that strong interaction effects due to pion exchange 
that arise in chiral effective field theory (EFT) renormalize the isovector coupling $a_1$.
The authors also present a new shell-model calculation for the Xenon isotopes 
$^{129}$Xe, $^{131}$Xe giving  numerical fits of the functions $S_{ij}$ that include
these new corrections.

Decomposing the spin-spin operator as a sum of the longitudinal and transverse electric operators projections
of the axial current,
see Appendix~\ref{App:A}, 
it is shown in Ref.~\cite{Menendez1} that the isovector coupling $a_1$ is renormalized as 
$\delta_{\mathcal{T}} a_1$ in $\mathcal{T^{\text{el 5}}}$ where
\begin{equation}
\delta_{\mathcal T} =1-2\frac{q^2}{\Lambda} +\delta a_1,
\label{deltaT}
\end{equation}
and $\delta_{\mathcal{L}}a_1$ in $\mathcal{L}^{5}$ with
\begin{equation}
\delta_{\mathcal L} =1-\frac{2g_{\pi pn}F_{\pi}q^2}{2m_N g_A (m^2_\pi +q^2)}
-\frac{2c_3 \rho q^2}{F^2_\pi (4m^2_\pi +q^2)}+\delta a_1.
\label{deltaL}
\end{equation}
The first $q^2$-dependent term of $\delta_{\mathcal T}$ and $\delta_{\mathcal L}$ 
arises at the one-body level (1b) currents, while the second $q^2$ dependent term in $\delta_{\mathcal L}$
and the momentum independent term 
\begin{equation}
\delta a_1 = -\frac{\rho}{F^2_\pi} I (\frac{1}{3} k +\frac{1}{6m_N})
\label{Eq:deltaa1}
\end{equation}
are due to  two-body (2b) currents.
Here $F_\pi =92.4$ MeV is the pion decay constant, $m_\pi =138.04$ MeV the pion mass,
$m_N$ the nucleon mass, $g_{\pi pn}=13.05$ the pion-nucleon coupling, $\Lambda =1040$ MeV
a cut-off scale, $\rho$ the nuclear
density, $\rho\in[0.10,0.12]$ fm$^{-3}$, $I\simeq 0.58-0.60$ is a dimensionless factor. 
The largest uncertainties are in the parameters
$c_3\in[-2.2,-4.78] $ GeV$^{-1}$ and $k=(2c_3-c_4) \in[7.2,14.0]$ GeV$^{-1}$, see Ref.~\cite{Menendez2}.

In principle, the calculation should be redone for all
the nuclei of experimental interest. However we now show how to include the new effects
in the existing calculations. The procedure is the following:
the corrections can be taken into account in the standard expression 
of the cross section, Eq.~(\ref{Eq:dsigmadE}), using the corrected static spin values 
$\langle\mathbf{S}_{p,n}\rangle_{\text{2b}}$ in $\sigma_A (0)$, Eq.~(\ref{Eq:sigmatot}), 
and using the known $S_{ii}$ function in the form factor (\ref{Eq:FFSD}) normalizing them to 1
at $q=0$ by factoring out their zero momentum transfer value.
We discuss it in two steps and then give the formula to obtain upper limits.

\subsection*{Comparison of normalized functions at q$\neq$ 0}

At finite momentum transfer there are two energy scales to be compared. 
Experiments that measure the recoil energy typically restrict the 
search window below 100 keV to reduce the background. Using the oscillator size 
parameter~\cite{Menendez1,Fitzpatrick1,resselldean} 
\[
b=\sqrt{\frac{41.467}{45A^{-1/3}-25A^{-2/3}}}\;\text{fm},
\]
100 keV corresponds to $y\sim 0.07$ in the case of $^{19}$F, $y\sim 0.39$ for $^{73}$Ge 
and $y\sim 0.83$ in the case of $^{131}$Xe. 
The other scale is set by the momentum transfer of the order of the pion mass, 100 MeV,
that corresponds to $y \sim 0.2$, 0.28 and 0.38, respectively. $q^2$ dependent corrections
can thus be neglected for light nuclei but should be considered in medium-heavy and heavy nuclei.

The normalized isospin functions have the property, as discussed in details in Refs.~\cite{divari,Cannoni1,Cannoni2},
\begin{equation}
\Phi_{00}(q)=\frac{S_{00} (q)}{S_{00} (0)}\simeq
\Phi_{11}(q)=\frac{S_{11} (q)}{S_{11} (0)}.
\label{Eq:isospinFF}
\end{equation}
This fact is  consequence of the isospin symmetry that treats proton and neutrons
on an equal footing. At small $y$, once normalized the isospin form factors behave like
$1-ay+by^2+{\mathcal{O}(y^3)} $. In light nuclei as $^{19}$F, $^{23}$Na, $^{29}$Si,
the polynomials are of low order, typically up to fourth degree, thus this relations 
is respected on a larger range of $y$, that anyway cover the range of momentum transfer
caused by dark matter elastic scattering. For medium heavy and heavy nuclei the more 
complicated structure requires higher order polynomials, from sixth to nine-th degree,
thus the differences are  more accentuated. 

In Fig.~\ref{Fig:1} we see that in the case of $^{129}$Xe Eq.~(\ref{Eq:isospinFF}) 
is verified  to a good approximation also at the level of the (2b)
functions of Ref.~\cite{Menendez1}, much less in the case of $^{131}$Xe (for unclear reasons).

Anyway, the most relevant thing to note in Fig.~\ref{Fig:1} is that
\begin{equation}
\frac{S_{11}^{\text{(1b)}}(q)}{S_{11}^{\text{(1b)}}(0)} 
\simeq \frac{S^{\text{(1b+2b)}}_{11}(q)}{S^{\text{(1b+2b)}}_{11}(0)},
\label{Eq:F11}
\end{equation}
for both $^{129}$Xe and $^{131}$Xe.
The correction $2/\Lambda^2$ is practically negligible while the 1b correction in $a_{\mathcal{T}}$ 
is already included in the standard calculations~\cite{engel,engelrev} in the form $1/(m_{\pi}^2 +q^2)$ using the 
Goldberger-Treiman relation $g_{\pi NN}\simeq g_A m_N/F_{\pi}$.
Thus, once normalized, the (1b) and (1b+2b) functions behave in the same way. 
\begin{figure}[t!]
\includegraphics*[scale=0.5]{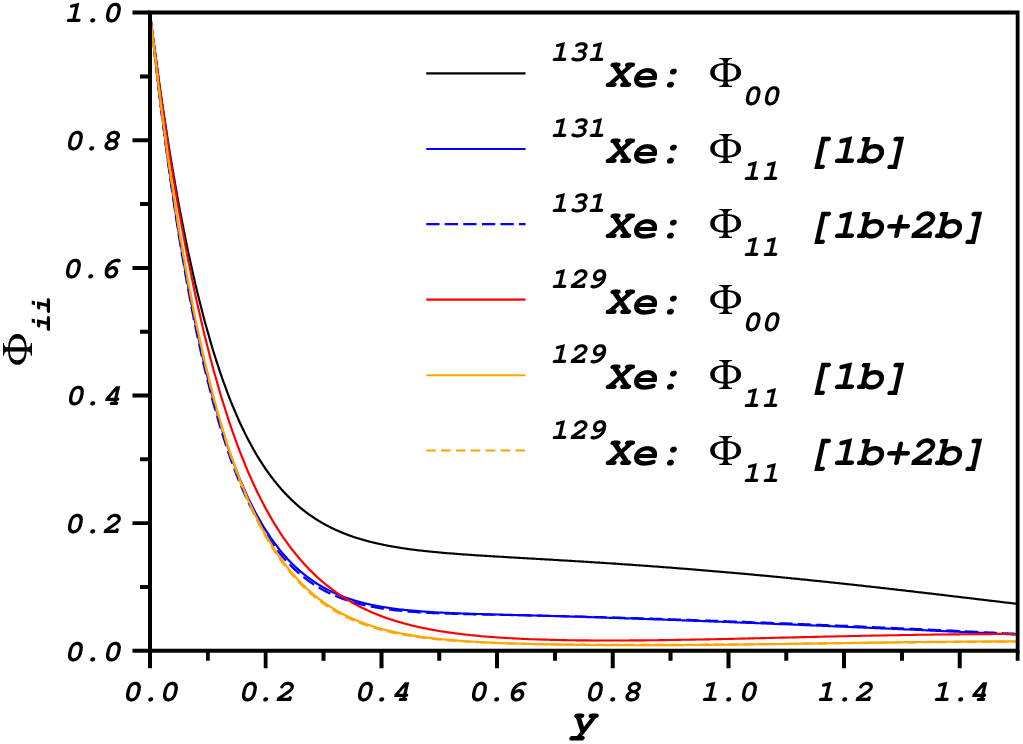}
\caption{Isospin form factors for the isotopes $^{129}$Xe and $^{131}$Xe built with the functions of 
Ref.~\cite{Menendez1}. See Eq.~(\ref{Eq:isospinFF}) for definitions.}
\label{Fig:3}
\end{figure}

\subsection*{Redefinition of the cross section at q=0}

At $q=0$ only the momentum independent 2-body currents effect $\delta a_1$ remains in Eqs.~(\ref{deltaT}), 
(\ref{deltaL}). In the following we set $\delta a_1 \equiv\delta$ to simplify the notation.
By replacing $a_1$ with $a_1(1+\delta)$ in $S(0)$
and using the defining relations of $a_{0,1}$ in terms of $a_{p,n}$
we obtain
\[
S(0)_{\text{2b}}=\frac{2J+1}{\pi}\frac{J+1}{J}|{a_p}\langle \mathbf{S}_{p}\rangle_{\text{2b}}
+a_n\langle \mathbf{S}_{n} \rangle_{\text{2b}}|^2,
\]
where we have defined
\begin{eqnarray}
\langle \mathbf{S}_{p} \rangle_{\text{2b}}&=& \langle \mathbf{S}_{p} \rangle
+\frac{\delta }{2}(\langle \mathbf{S}_{p} \rangle - \langle \mathbf{S}_{n} \rangle),\crcr
\langle \mathbf{S}_{n} \rangle_{\text{2b}}&=& \langle \mathbf{S}_{n} \rangle
-\frac{\delta }{2}(\langle \mathbf{S}_{p} \rangle - \langle \mathbf{S}_{n} \rangle).
\label{Eq:SpSn2b}
\end{eqnarray}
Obviously the total contribution of the proton and neutron groups to the total angular
momentum does not change, 
$\langle \mathbf{S}_{p} \rangle_{\text{2b}}+\langle \mathbf{S}_{n} \rangle_{\text{2b}}
=\langle \mathbf{S}_{p} \rangle + \langle \mathbf{S}_{n} \rangle$.
Note that also the sign of the spin matrix elements is important in the determination 
of the size of the correction. 

Considering the possible range in $\rho$, $I$ and $k$ we find for the interval and the average value:
\begin{equation}
\delta  \in [-0.135,-0.314],\;\;\text{and}\;\; \delta_{\text{av}}=-0.224. 
\label{Eq:deltanum}
\end{equation}
In Table~\ref{Tab:1} we report the spin values for the most important isotopes employed
as detecting medium in actual experiments and their values after the inclusion of the 2-body
current effects using the average value  $\delta_{\text{av}}$. The starting values are taken from
papers that claim to present the best shell-model calculation for the given isotope
(we rounded to 3 decimal digits where necessary).  
\begin{table}[b!]
\caption{Spin matrix elements for the odd-mass isotopes employed in dark matter search experiments.
The values of the last two columns include the 2-body currents corrections as given in Eq.~(\ref{Eq:SpSn2b})
with $\delta=\delta_{\text{av}}$ of Eq.~(\ref{Eq:deltanum}).}
\label{Tab:1}
\begin{ruledtabular}
\begin{tabular}{cccccc}
 Isotope (J) [Ref.] & $\langle \mathbf{S}_{p} \rangle$ & $\langle \mathbf{S}_{n} \rangle$ & $\langle \mathbf{S}_{p} 
 \rangle_{\text{2b}}$ & $\langle \mathbf{S}_{n} \rangle_{\text{2b}}$ \\	
\hline
\\
 $^{19}$F  (1/2)~\cite{divari}  & 0.475     & -0.009  & 0.421  & 0.045 &\\
\\
$^{23}$Na (3/2)~\cite{divari} & 0.248    & 0.020   & 0.222   & 0.045 &\\
\\
$^{29}$Si (1/2)~\cite{divari} & -0.002   & 0.133   & 0.013   & 0.118 &\\
\\
$^{73}$Ge (9/2)~\cite{dimitrov} & 0.030     & 0.378    & 0.069   & 0.339 &\\
\\
$^{127}$I (5/2)~\cite{resselldean}  & 0.309    & 0.075    & 0.283   & 0.101 &\\
\\
$^{129}$Xe (1/2)~\cite{Menendez1}  & 0.010   & 0.329   &  0.046  & 0.293  &\\
\\
$^{131}$Xe (3/2)~\cite{Menendez1} & -0.009  & -0.272  & -0.038  & -0.242  &\\
\\
$^{133}$Cs (7/2)~\cite{toivanen}  & -0.318  & 0.021   & -0.280   & -0.017 &
\end{tabular}
\end{ruledtabular}
\end{table}

The case of $^{19}$F needs a particular comment. The experiments employing $^{19}$F,
COUPP~\cite{coupp},
SIMPLE~\cite{simple} and PICASSO~\cite{picasso}, 
employ the spin values of Ref.~\cite{pacheco}, often quoting the compilation of Ref.~\cite{gondolo},
$\langle \mathbf{S}_{p} \rangle = 0.441$, $\langle \mathbf{S}_{n} \rangle =-0.109 $.
The successive refined shell-model calculation of Ref.~\cite{divari} using the more realistic Wildhental 
interaction found $\langle \mathbf{S}_{p} \rangle = 0.4751$ and $\langle \mathbf{S}_{n} \rangle = -0.0087$.
The  protons contribution is thus similar but the neutrons contribution
is now a factor $\simeq 50$ smaller and not 4 times smaller than the protons one.

The new calculations of Ref.~\cite{Fitzpatrick1} allows us to clarify the question.
These authors use the same Wildhental interaction, the harmonic oscillator wave functions 
and the full $s-d$ shell structure, thus the results should agree with the ones of Ref.~\cite{divari}.
They furnish the analytical expressions of the response
functions for $^{19}$F but do not calculate the static spin values. 
We use Eq.~(\ref{Eq:Sii}) to obtain $S_{ii}(0)$ and Eqs.~(\ref{Eq:coeff0}),~(\ref{Eq:+SpSn})
to get the spin values.
We find $\langle \mathbf{S}_{p} \rangle = 0.475(5)$ and $\langle \mathbf{S}_{n} \rangle = -0.008(68)$,
in perfect numerical agreement with the values of Ref.~\cite{divari}. 
Such an agreement is also found in the case of $^{23}$Na, see Fig.~\ref{Fig:6} in Appendix~\ref{App:B}. 
In Ref.~\cite{Fitzpatrick1} new shell-model calculations for $^{73}$Ge, $^{127}$I,
$^{129,131}$Xe are also given. We have verified that in these cases the spin values differ 
largely from the other known calculations, see also Ref.~\cite{Fitzpatrick2}.
Given that the same authors in Ref.~\cite{Fitzpatrick1} state explicitly that these calculations
are "exploratory", we do not use them in Table~\ref{Tab:1}.

\subsection*{Final formula}

Suppose that for a certain nucleus we have the structure functions at the (2b) level
as in Ref.~\cite{Menendez1}, then
the differential cross section when the protons or neutrons coupling dominate is
\begin{equation}
\frac{d\sigma}{dE_R}  =\frac{m_A}{2 \mu^2_A v^2 } \frac{\mu_A^2}{\mu_p^2}\frac{4}{3} \frac{\pi}{2J+1} 
\sigma_{{p,n}} S^{(\text{1b+2b})}_{{p,n}} (q),
\label{eq:ds_ex}
\end{equation}
where $\mu_A$ and $\mu_p$ are the WIMP-nucleus and WIMP-nucleon reduced masses.
The neutron and proton group structure functions are
\begin{eqnarray}
S^{(\text{1b+2b})}_{p,n} (q)&=&S^{(\text{1b})}_{00}(q)+S^{(\text{1b+2b})}_{11}(q)\pm S^{(\text{1b+2b})}_{01}(q)
\nonumber\\
              &\equiv&\left(\sqrt{S^{(\text{1b})}_{00}(q)}\pm\sqrt{S^{(\text{1b+2b})}_{11}(q)}\right)^2,
\label{eq:Spn}
\end{eqnarray}
where the equivalence follows from the discussion of Section~\ref{Sec:1} and Appendix~\ref{App:A}.
If we want to use a normalized form factor, then Eq.~(\ref{eq:ds_ex}) is substituted by
\begin{equation}
\frac{d\sigma}{dE_R}  =\frac{m_A}{2 \mu^2_A v^2 } \frac{\mu_A^2}{\mu_p^2}\frac{4}{3} \frac{J+1}{J}\sigma_{p,n}
\langle\mathbf{S}_{p,n}\rangle^2_{\text{2b}}\frac{S^{(\text{1b+2b})}_{p,n} (q)}{S^{(\text{1b+2b})}_{p,n}(0)}.
\label{eq:ds_ex_ff}
\end{equation}
Eqs.~(\ref{eq:ds_ex}) and (\ref{eq:ds_ex_ff}) are mathematically equivalent.
\begin{figure}[t!]
\begin{center}
\includegraphics*[scale=0.48]{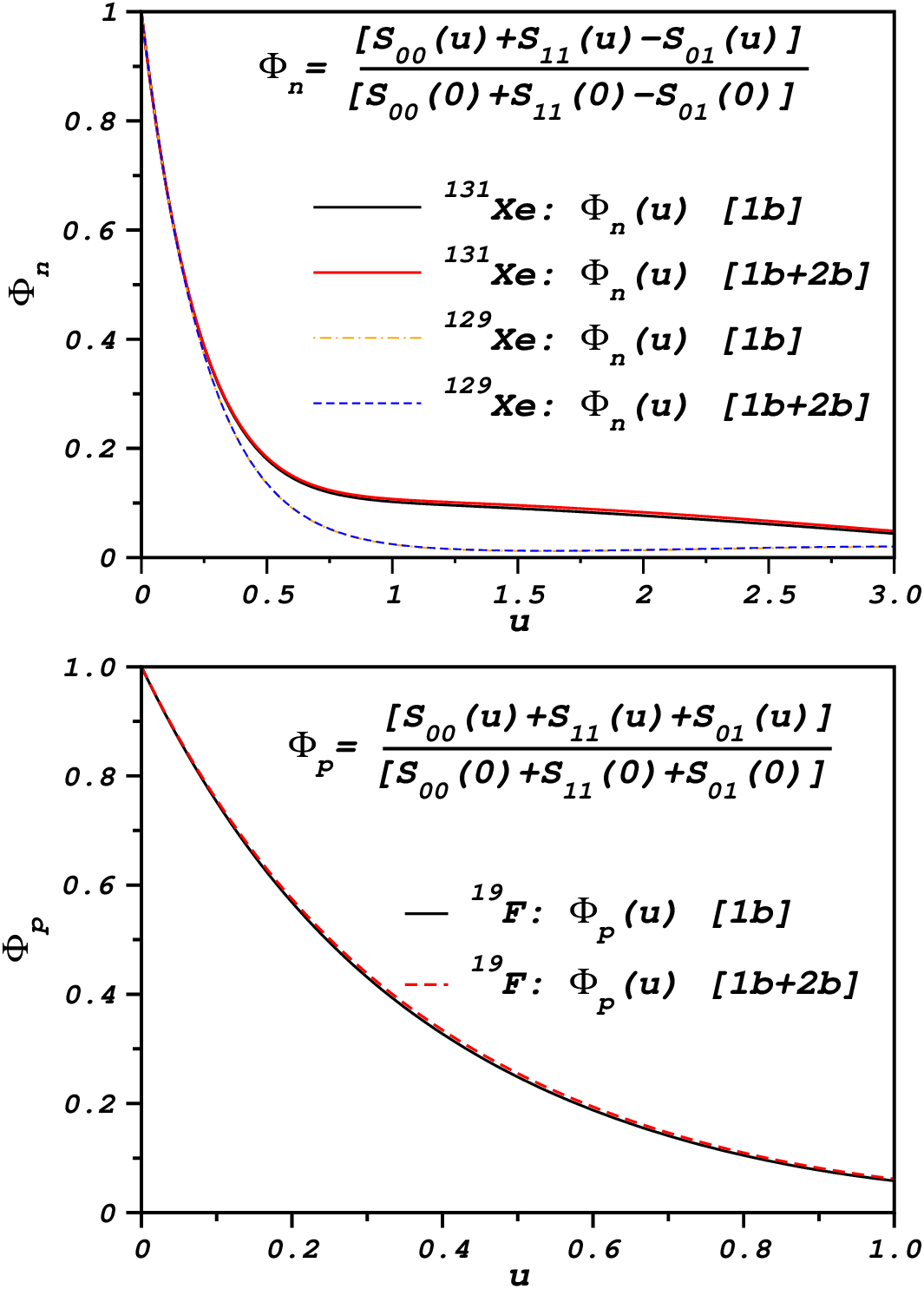}
\caption{Top panel: normalized neutrons form factor for the isotopes $^{129}$Xe and $^{131}$Xe built 
with the functions of Ref.~\cite{Menendez1}. Bottom panel: normalized protons form factor for $^{19}$F
built with the functions of Ref.~\cite{vergados_nuovo}. }
\label{Fig:4}
\end{center}
\end{figure}

In Fig.~\ref{Fig:4} we show the normalized neutrons form factor 
using the (1b) and (1b+2b) functions of Ref.~\cite{Menendez1} for the Xenon isotopes
and the normalized protons form factor of $^{19}$F
with the new functions of Ref.~\cite{vergados_nuovo}. We see that, as expected,
$\Phi^{(\text{1b})}_{p,n} =\Phi^{(\text{1b+2b})}_{p,n}$. 

It is clear then that the following expression for the cross section 
\begin{equation}
\frac{d\sigma}{dE_R}  =\frac{m_A}{2 \mu^2_A v^2 } \frac{\mu_A^2}{\mu_p^2}\frac{4}{3} \frac{J+1}{J}\sigma_{p,n}
\langle\mathbf{S}_{p,n}\rangle^2_{\text{2b}}\frac{S^{(\text{1b})}_{p,n} (q)}{S^{(\text{1b})}_{p,n}(0)},
\label{eq:ds_ex_1b}
\end{equation} 
will give the same limits obtained with Eq.~(\ref{eq:ds_ex}) or Eq.~(\ref{eq:ds_ex_ff}).

\section{XENON100 and COUPP limits}
\label{Sec:3}

The XENON100 collaboration has recently published~\cite{xenon100_225_SD} the most stringent
exclusion limits on the WIMP-neutron cross section using the new updated functions of Ref.~\cite{Menendez1}.
On the other hand, for example, the best limits on the WIMP-proton cross section
come from experiments using $^{19}$F, in particular from COUPP~\cite{coupp}.

We calculate the 90$\%$ confidence level upper limit on the WIMP-neutron cross section
using the method of Ref.~\cite{Cannoni2} with the standard  Maxwellian distribution truncated
at $v_{esc}=544$ km/s, the velocity of the Sun $v_0=230$ km/s,
the velocity of the Earth $v_E=244$ km/s and  $\rho_0 =0.3$ GeV/cm$^{3}$ 
for the local dark matter density.

\begin{figure}[tb!]
\includegraphics*[scale=0.36]{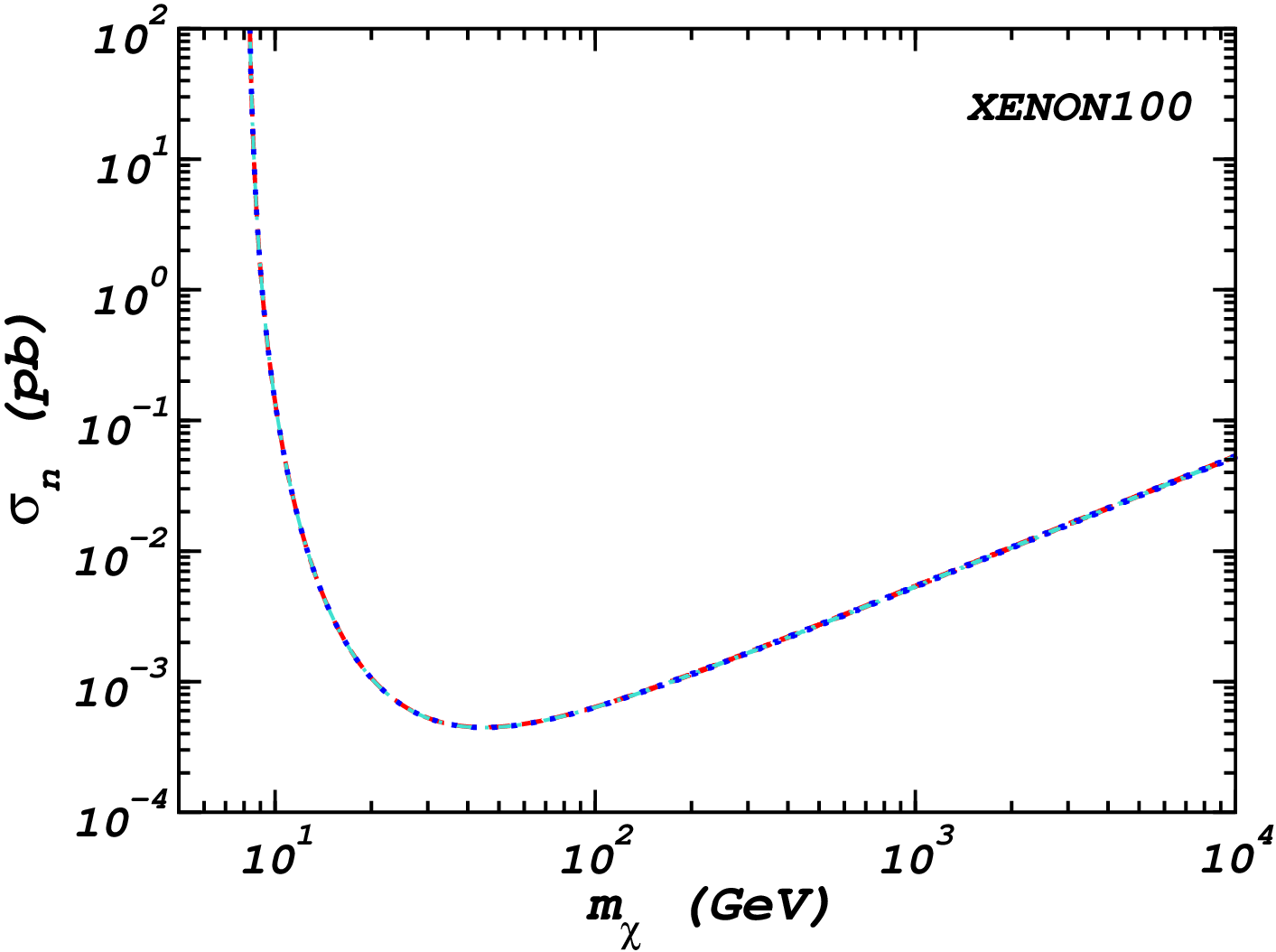}
\includegraphics*[scale=0.35]{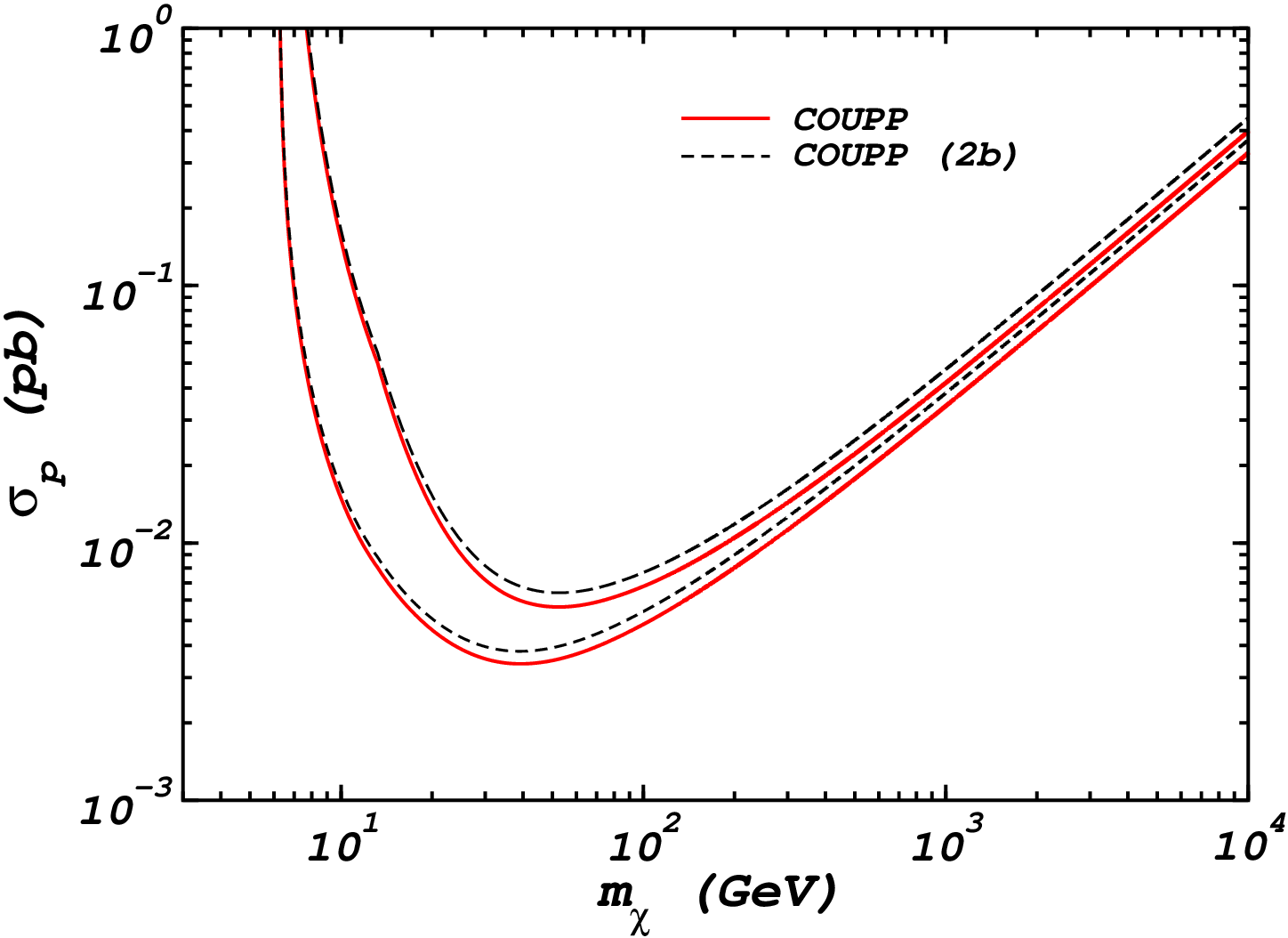}
\caption{Top panel: 90 $\%$ confidence level upper limit on the WIMP-neutron cross section obtained 
with the last data of XENON100. The curve $(a)$ is calculated with Eq.~(\ref{eq:ds_ex}), the curve $(b)$
with Eq.~(\ref{eq:ds_ex_ff}) and the curve $(c)$ with Eq.~(\ref{eq:ds_ex_1b}).
Bottom panel: The solid red lines are the  COUPP 90\% confidence level 
corresponding to two different models for the bubble nucleation efficiency of fluorine~\cite{coupp}.
The black dashed lines are calculated with the corrected spin values of Table~\ref{Tab:1}.
See the text for further details on how the limits in both panels are calculated.}
\label{Fig:5}
\end{figure}

We use the exposure 34$\times$224.6 kg$\times$year and the energy dependent acceptance
fitting the solid blue curve of Fig.~1 in Ref.~\cite{xenon100_225_SI} with a six degree polynomial. 
In the fiducial energy range (6.6-30.5) keV, 
2 observed events and 1 expected mean background imply an 90 $\%$ confidence level upper limit of 4.9
according to the  Feldman and Cousins method.
The limits calculated with Eqs.~(\ref{eq:ds_ex}), (\ref{eq:ds_ex_ff}) and (\ref{eq:ds_ex_1b}) are shown 
in the top panel of Fig.~\ref{Fig:5}: as expected they are identical.
Note also that our simple method gives the same curve of Ref.~\cite{xenon100_225_SD} except for the low mass region
where the our curve is above a mass of 8 GeV while the limit of Ref.~\cite{xenon100_225_SD} starts to constrain 
masses above 6 GeV.

In Fig.~\ref{Fig:5}, bottom panel, we show the impact of the (2b) currents on the limits 
of COUPP~\cite{coupp} (CF$_3$I). 
The efficiencies, exposures and a threshold 
are specified in Ref.~\cite{coupp}. The two red curves correspond
to  two different models for the bubble nucleation efficiency of fluorine. 
Note that we find the agreement with the published curves 
using an upper limit on the number of events of $N^{UL}=14.5$, that is the $90\%$ confidence 
level limit obtained with the Feldman-Cousins method with 13 observed events 
(7 of 20 nuclear recoil events removed by the time isolation cut) and an expected background
of 4.5.
The black dashed lines are the new limits calculated with $\delta_{\text{av}}$. 
The reduction in the mass region above the minimum is about 10\%.

These examples clearly show that one can safely use Eq.~(\ref{eq:ds_ex_1b}) to set upper 
limits on $\sigma_{p,n}$ even if
for the nucleus at hand the functions $S^{(\text{1b+2b})}_{ij}$ have not been published,
but only the standard functions, that in the language of Ref.~\cite{Menendez1}
are $S^{(\text{1b})}_{ij}$, are available. 

Our results allow also to easily include
the effects in analysis concerning the impact
of nuclear physics uncertainties on the reconstruction of the WIMP properties,
mass and cross section,
from the experiments with spin-dependent sensitivity~\cite{cerdeno}.

\section{Summary}
\label{Sec:4}

We here summarize the main results of the paper, that in formulas are
given by Eqs.~(\ref{Eq:F11}),~(\ref{Eq:SpSn2b}),~(\ref{Eq:deltanum}),~(\ref{eq:Spn}), (\ref{eq:ds_ex_1b}): 
\begin{itemize}
\item
Among the new corrections to the isovector couplings arising in chiral effective field theory
as found in Ref.~\cite{Menendez1} the only relevant one, as far as concerns the recoil energy range
of interest for direct detection experiments, is the momentum independent $\delta a_1$, Eq.~(\ref{Eq:deltaa1}).
The values are in the interval given in Eq.~(\ref{Eq:deltanum}) due to uncertainties in the knowledge of
the effective field theory parameters entering in Eq.~(\ref{Eq:deltaa1}).
\item
The correction $\delta a_1$ affects the total WIMP-nucleus cross section at $q=0$. 
In the standard proton-neutron representation
it induces a WIMP-protons coupling even in the case that for a given nucleus the spin is determined by the 
neutrons group (and vice versa), as shown by Eq.~(\ref{Eq:SpSn2b}). 
\item
By formally redefining the spin matrix elements
as in Eq.~(\ref{Eq:SpSn2b}) and using the normalized form factor with  
the standard isospin functions, Eq.~(\ref{eq:ds_ex_1b}) we get the same cross section and the same upper limits
obtained using Eq.~(\ref{eq:ds_ex}). Fig.~\ref{Fig:5} shows it explicitly and unambiguously in the case of the 
WIMP-neutron cross section based on the last XENON100 data.
\item
The standard formalism for treating spin-dependent scattering, even including the two-body currents effects 
at the level of Ref.~\cite{Menendez1}, is redundant because the isospin interference function $S_{01}$ is not
an independent function, as believed for more than 20 years, 
but simply the double product of the square root of the other two.
\item
Experiments using $^{19}$F should use the spin matrix elements of Ref.~\cite{divari} that are obtained
with shell-model calculation employing the Wildenthal potential and not the old values of Ref.~\cite{pacheco}.
We have shown that the new calculation of Ref.~\cite{Fitzpatrick1} agrees with the results of Ref.~\cite{divari}.

\end{itemize} 

\section*{Acknowledgments}
Work supported by MultiDark under Grant No. CSD2009-00064 of the
Spanish MICINN  Consolider-Ingenio 2010 Program. Further support is provided by the
MICINN project FPA2011-23781 and from the Grant MICINN-INFN(PG21)AIC-D-2011-0724. 

\appendix

\section{The isoscalar-isovector interference function}
\label{App:A}

The spin-spin operator  
\begin{equation}
{O}(q)=4 {\mathbf{S}}_{\chi} \cdot \sum_{i=1}^{A}\frac{1}{2}(a_0\openone  +a_1 {\tau}^3_i ) 
{\mathbf{S}}_{i}
e^{-i\mathbf{q}\cdot \mathbf{r}_i}
\end{equation}
with ${\mathbf{S}}_{i}$ and $\mathbf{r}_i$ the spin and coordinates of the $i$--th nucleon
${\tau}_3 |p\rangle =|p\rangle$, ${\tau}_3 |n\rangle =-|n\rangle$ and $\openone$,
the identity operator in isospin space, is expressed in the standard formalism~\cite{engel} 
(also in Ref.~\cite{Fitzpatrick1}) as the sum of the transverse electric 
and longitudinal projections of the axial current
\[
{O}(q)=4 \hat{\mathbf{S}}_{\chi} \cdot (\mathcal{T}^{el\,5}(q)\hat{e}_{\pm} + \mathcal{L}^{5}(q)\hat{e}_0).
\]
$\hat{e_0}$ is a spherical unit vector in the direction of the quantization axis taken to be one of 
$\mathbf{q}$ and $\hat{e}_{\pm}$ are the unit vector orthogonal to this direction.   
These operators are decomposed into multipoles as
\begin{eqnarray}
\mathcal{T}^{el\,5}&=& \sum\limits_{i=1}^{A}\frac{1}{2}(a_0\openone  +a_1 \delta_{\mathcal{T}} \hat{\tau}^3_i )
T(q),\\
\mathcal{L}^{5}&=&\sum\limits_{i=1}^{A}\frac{1}{2}(a_0\openone  +a_1 \delta_{\mathcal{L}} \hat{\tau}^3_i )L(q),
\end{eqnarray}
where the detailed expressions for $T(q)$ and $L(q)$ are unnecessary for our argument.  
The coefficients $\delta_{\mathcal{T}}$ and $ \delta_{\mathcal{L}}$ account for the effects 
discussed in Section~\ref{Sec:2}.
Given the nuclear wave function $|A\rangle$, then
\begin{equation}
|\mathcal{M}|^2 \propto (|\langle A|| \mathcal{T}^{el\,5}(q)||A\rangle|^2 +
 |\langle A|| \mathcal{L}^{5}(q) ||A\rangle|^2) .
\label{Msquared}
\end{equation}
There is no $\mathcal{T}^{el\,5}-\mathcal{L}^{5}$ interference in the modulus squared 
because
of the orthogonality, but each term in Eq.~(\ref{Msquared}) gives contributions proportional 
to $a^2_0$, $a^2_1$ and to the interference $a_0 a_1$. 
The functions $S_{ij}(q)$ arise as a sum of pieces after a rearrangement of these terms:
\begin{equation}
|\mathcal{M}|^2 \propto (a^2_0 S_{00}(q)+a^2_1 S_{11}(q) +a_0 a_1 S_{01}(q)).
\label{Msquared1}
\end{equation}

\begin{figure}[t!]
\includegraphics*[scale=0.59]{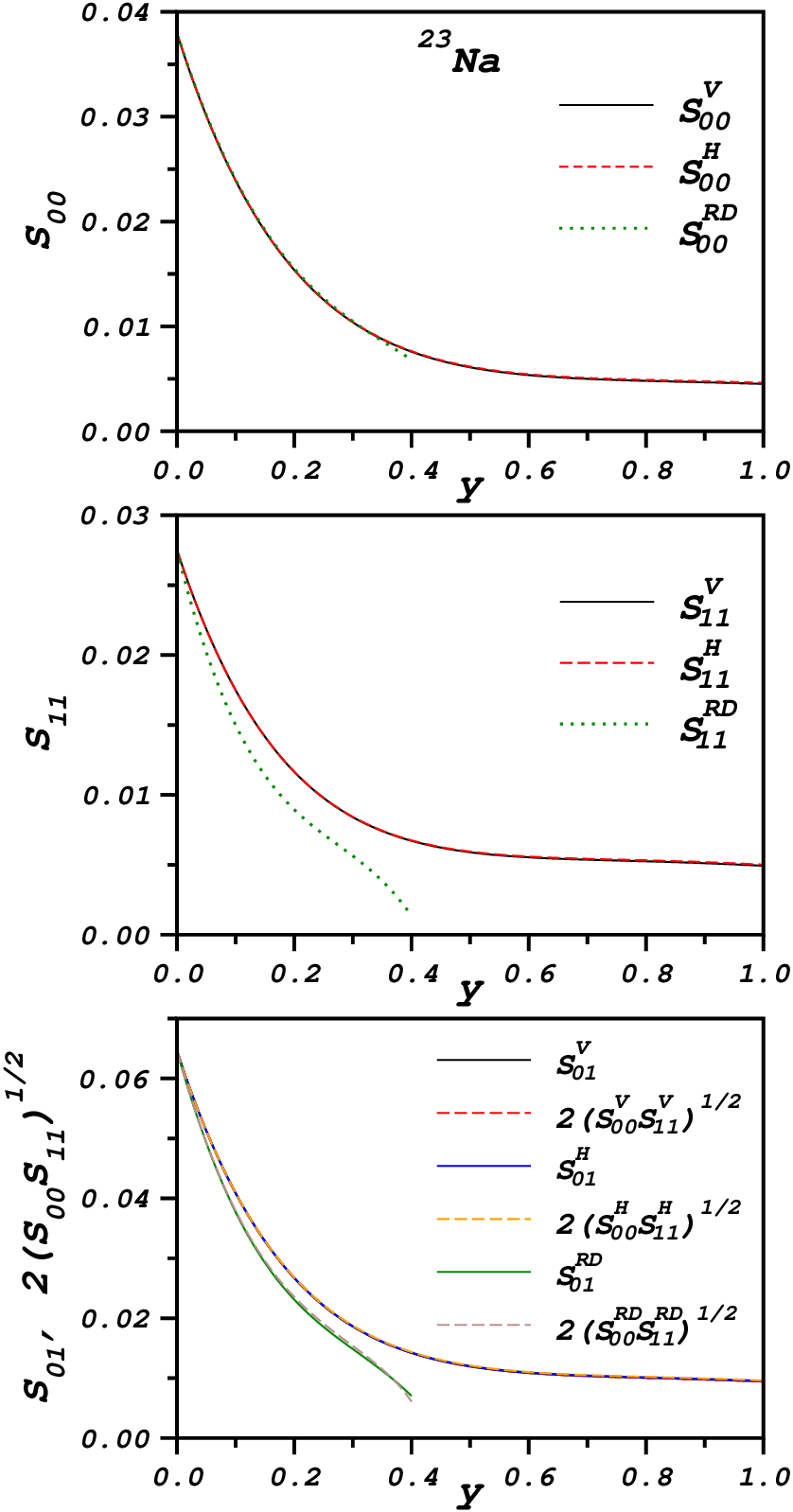}
\caption{The structure functions for $^{23}$Na. The superscript $V$ and $H$ indicate the
that the functions are built from the functions of Ref.~\cite{divari} and of Ref.~\cite{Fitzpatrick1}
respectively, using the prescriptions of Appendix~\ref{App:B}. The superscript $RD$ indicate 
the polynomial expressions of Ref.~\cite{resselldean}. Note that the four curves corresponding to
$S^V_{ij}$ and $S^H_{ij}$ are almost identical and practically undistinguishable. }
\label{Fig:6}
\end{figure}

On the other hand, equivalently,
the spin-spin operator can be decomposed,  as done in Refs.~\cite{divari}, 
\[
{O}(q)=4 {\mathbf{S}}_{\chi} \cdot (a_0 {\bm{S}}_0 +a_1 {\bm{S}}_1 ),
\]
where
\begin{eqnarray}
{\bm{S}}_0 (q)=\frac{1}{2}\sum\limits_{i=1}^{A} \openone
{\mathbf{S}}_i e^{-i\mathbf{q}\cdot \mathbf{r}_i},\;\;
{\bm{S}}_1 (q)=\frac{1}{2}\sum\limits_{i=1}^{A} \hat{\tau}_i^3 
{\mathbf{S}}_i e^{-i\mathbf{q}\cdot \mathbf{r}_i}.
\nonumber
\end{eqnarray}
are the nuclear spin operators associated with the isospin couplings.
Performing the multipole decomposition on these operators we are led to
\begin{eqnarray}
|\mathcal{M}|^2 \propto 
|a_0 \langle A|| {\bm{S}}_0 (q)||A\rangle + a_1 \delta_1 \langle A||{\bm{S}}_1  (q)||A\rangle |^2.
\label{Msquared2}
\end{eqnarray}
Now all the corrections are included in $\delta_1$ and we can define two functions
such that Eq.~(\ref{Msquared2}) takes the form 
\begin{equation}
|\mathcal{M}|^2 \propto (a_0 S^2_0(q) +a_1  S^2_1(q) +2a_0 a_1  S_0(q) S_1(q) ).
\label{Msquared3}
\end{equation}
Eq.~(\ref{Msquared1}) and Eq.~(\ref{Msquared3}) are equivalent:
the isoscalar and the isovector functions $S_{00}$ and $S_{11}$ 
are squared of some functions $S_0 (q)$ and $S_1 (q)$
and the interference function is the double product
$ 2 {S_{0}(q) S_{1}(q)}$.

\section{Three equivalent notations}
\label{App:B}

In the notation of Ref.~\cite{divari} we can identify
$S_0=\Omega_0$, $S_1 =\Omega_1$. From these the normalized form factors are defined,
$F_{ij}(q)=\Omega^2_{ij}(q)/\Omega^2_{ij}(0)$. 

In the notation of Ref.~\cite{Fitzpatrick1}  
the basic spin response functions are, $N,N'=p,n$, $F^{N,N'}_{\Sigma'}$ 
(corresponding to the axial transverse electric operator $\mathcal{T}^{\text{el 5}}$) 
and $ F^{N,N'}_{\Sigma''}$ (corresponding to the axial longitudinal operator $\mathcal{L}^5$), thus
\begin{equation}
F^{N,N'}_{44}=\frac{1}{16}(F^{N,N'}_{\Sigma'} + F^{N,N'}_{\Sigma''}).
\label{Eq:Fsigmas}
\end{equation}
The subscript "44" designate the spin-spin interaction.
From these, the form factors in the isospin representation are  
\begin{eqnarray}
F^{00}_{44}&=&\frac{1}{4}(F^{p,p}_{44}+F^{n,n}_{44}+2F^{p,n}_{44}),\crcr
F^{11}_{44}&=&\frac{1}{4}(F^{p,p}_{44}+F^{n,n}_{44}-2F^{p,n}_{44}).
\label{Eq:F44}
\end{eqnarray}

The standard isoscalar and isovector functions $S_{00}$ and $S_{11}$ read:
\begin{equation}
S_{ii}(q)\equiv \frac{2J+1}{16\pi}\Omega^2_{i}(0)F_{ii}(q) \equiv 4\frac{2J+1}{\pi}F^{ii}_{44}(q).
\label{Eq:Sii}
\end{equation}

This equivalence is illustrated in Fig.~\ref{Fig:6} for the isotopes $^{23}$Na, that to our knowledge
is the only nucleus for which the calculation in the three formalisms exists in literature. The results of 
Ref.~\cite{divari} and Ref.~\cite{Fitzpatrick1} coincide exactly for both $S_{00}$ and $S_{11}$.
The isoscalar function of Ref.~\cite{resselldean} agrees with the others two, while it is not the case for 
the isovector function. Note that these last ones are furnished as polynomial fits that 
are valid up to $y\sim 0.35 $, for $y>0.4$ they become negative. 
The double product rule is respected by the three calculations.


\begin{thebibliography}{100}

\bibitem{engel}
J.~Engel,
Phys.\ Lett.\ B {\bf 264}, 114 (1991).


\bibitem{engelrev}
J.~Engel, S.~Pittel and P.~Vogel,
Int.\ J.\ Mod.\ Phys.\  E {\bf 1}, 1 (1992).

\bibitem{jungman} 
G.~Jungman, M.~Kamionkowski and K.~Griest,
Phys.\ Rept.\  {\bf 267}, 195 (1996)
[hep-ph/9506380].

\bibitem{goodman} 
M.~W.~Goodman and E.~Witten,
Phys.\ Rev.\ D {\bf 31}, 3059 (1985).

\bibitem{kurylov} 
A.~Kurylov and M.~Kamionkowski,
Phys.\ Rev.\ D {\bf 69}, 063503 (2004)
[hep-ph/0307185].

\bibitem{Fitzpatrick1} 
A.~L.~Fitzpatrick, W.~Haxton, E.~Katz, N.~Lubbers and Y.~Xu,
JCAP {\bf 1302}, 004 (2013)
[1203.3542].



\bibitem{Menendez1} 
J.~Menendez, D.~Gazit and A.~Schwenk,
Phys.\ Rev.\ D {\bf 86}, 103511 (2012)
[1208.1094].


\bibitem{xenon100_225_SD} 
E.~Aprile {\it et al.}    [XENON100 Collaboration],
arXiv:1301.6620 [astro-ph.CO]

\bibitem{xenon100_225_SI} 
E.~Aprile {\it et al.}  [XENON100 Collaboration],
Phys.\ Rev.\ Lett.\  {\bf 109}, 181301 (2012)
[1207.5988].

\bibitem{coupp} 
E.~Behnke {\it et al.},  [COUPP Collaboration],
Phys.\ Rev.\ D {\bf 86}, 052001 (2012)
[1204.3094].

\bibitem{mica} 
J.~Engel, M.~T.~Ressell, I.~S.~Towner and W.~E.~Ormand,
Phys.\ Rev.\ C {\bf 52}, 2216 (1995)
[hep-ph/9504322].

\bibitem{dimitrov}
V.~I.~Dimitrov, J.~Engel and S.~Pittel,
Phys.\ Rev.\  D {\bf 51}, 291 (1995).
[hep-ph/9408246]

\bibitem{resselldean}
M.~T.~Ressell and D.~J.~Dean,
Phys.\ Rev.\  C {\bf 56}, 535 (1997).
[hep-ph/9702290]

\bibitem{Menendez2} 
J.~Menendez, D.~Gazit and A.~Schwenk,
Phys.\ Rev.\ Lett.\  {\bf 107}, 062501 (2011)
[1103.3622].

\bibitem{divari}
P.~C.~Divari, T.~S.~Kosmas, J.~D.~Vergados and L.~D.~Skouras,
Phys.\ Rev.\  C {\bf 61}, 054612 (2000).

\bibitem{Cannoni1}
M.~Cannoni, J.~D.~Vergados and M.~E.~Gomez,
Phys.\ Rev.\  D {\bf 83}, 075010 (2011).
[arXiv:1011.6108]

\bibitem{Cannoni2}
M.~Cannoni,
Phys.\ Rev.\ D {\bf 84}, 095017 (2011)
[1108.4337]. 

\bibitem{vergados_nuovo}
P.~C.~Divari and J.~D.~Vergados,
arXiv:1301.1457.

\bibitem{simple} 
M.~Felizardo
{\it et al.} [SIMPLE Collaboration],
Phys.\ Rev.\ Lett.\  {\bf 108}, 201302 (2012)
[1106.3014].

\bibitem{picasso} 
S.~Archambault {\it et al.},  [PICASSO Collaboration],
Phys.\ Lett.\ B {\bf 711}, 153 (2012)
[1202.1240].

\bibitem{pacheco}
A.~F.~Pacheco and D.~Strottman,
Phys.\ Rev.\ D {\bf 40}, 2131 (1989).

\bibitem{gondolo} 
D.~R.~Tovey, R.~J.~Gaitskell, P.~Gondolo, Y.~A.~Ramachers and L.~Roszkowski,
Phys.\ Lett.\ B {\bf 488}, 17 (2000)
[hep-ph/0005041].

\bibitem{toivanen}
P.~Toivanen, M.~Kortelainen, J.~Suhonen and J.~Toivanen,
Phys.\ Rev.\ C {\bf 79}, 044302 (2009).

\bibitem{Fitzpatrick2} 
A.~L.~Fitzpatrick, W.~Haxton, E.~Katz, N.~Lubbers and Y.~Xu,
arXiv:1211.2818.

\bibitem{cerdeno} 
D.~G.~Cerdeno, M.~Fornasa, J.~-H.~Huh and M.~Peiro,
Phys.\ Rev.\ D {\bf 87}, 023512 (2013)
[1208.6426].


\end{thebibliography}
\end{document}